\documentclass[journal,twoside,web]{ieeecolor}

\usepackage{generic}
\usepackage{cite}
\usepackage{amsmath,amssymb,mathtools}
\usepackage{graphicx}
\usepackage{bm}
\usepackage{dashrule}
\usepackage{hyperref}
\usepackage{xurl}
\usepackage[ruled,linesnumbered]{algorithm2e}
\usepackage{textcomp}

\newtheorem{theorem}{Theorem}
\newtheorem{proposition}{Proposition}
\newtheorem{corollary}{Corollary}

\newtheorem{assumption}{Assumption}
\newtheorem{definition}{Definition}
\newtheorem{remark}{Remark}

\newcommand{\R}{\mathbb{R}}
\newcommand{\C}{\mathbb{C}}
\newcommand{\Z}{\mathbb{Z}}
\newcommand{\E}{\mathbb{E}}

\newcommand{\Sym}{\mathbb{S}}
\newcommand{\Prob}{\mathbb{P}}

\newcommand{\norm}[1]{\left\lVert #1 \right\rVert}

\newcommand{\col}{\operatorname{col}}

\newcommand{\eig}{\operatorname{eig}}

\renewcommand{\vec}{\rm{vec}}
\newcommand{\blkdiag}{\operatorname{blkdiag}}
\newcommand{\id}{\mathrm{id}}

\newenvironment{pfof}[1]{\vspace{1ex}\noindent{\itshape Proof of
    #1:}\hspace{0.5em}} {\hfill\oprocend\vspace{1ex}}

\newcommand\oprocendsymbol{\hbox{$\square$}}
\newcommand\oprocend{\relax\ifmmode\else\unskip\hfill\fi\oprocendsymbol}

\usepackage{xcolor}

\SetKwInOut{AlgData}{Data}
\SetKwInOut{AlgInput}{Input}
\SetKwInOut{AlgOutput}{Output}

\begin{document}

\title{Distributionally Robust Frequency Response Estimation Using Non-Steady-State Frequency-Domain Data}
\author{Stefan Damjanovic, Klaske van Heusden, \IEEEmembership{Member, IEEE}\, and John W. Simpson-Porco, \IEEEmembership{Senior Member, IEEE}
\thanks{This work was supported in part by NSERC Discovery Grant RGPIN-2024-05523 and by the Ontario Early Researcher Award.}
\thanks{Stefan Damjanovic and John W. Simpson-Porco are with the Department of Electrical and Computer Engineering, University of Toronto, 10 King's College Road, Toronto, ON, M5S 3G4, Canada (e-mail: stefan.damjanovic@mail.utoronto.ca, jw.simpson@utoronto.ca)}
\thanks{Klaske van Heusden is with University of British Columbia, Canada (e-mail: 
klaske.vanheusden@ubc.ca).}
}

\maketitle     
\begin{abstract}
This paper develops a distributionally robust optimization (DRO) framework for identifying the frequency response function (FRF) of an unknown linear time-invariant system from repeated noisy non-steady-state input-output measurements. Building on a recent frequency-domain extension of Willems' fundamental lemma, we formulate FRF estimation as a DRO problem, wherein robustness is desired with respect to the unknown probability distribution generating the noisy data. The proposed estimator computes FRF estimates by minimizing a convex upper bound on the DRO cost and allows for frequency-dependent calibration of the Wasserstein metric; the tractable upper bound consists of an empirical fit and a regularizer that penalizes the largest weighted sensitivity to spectral perturbations. Under suitable assumptions, we prove that the FRF estimation error is upper bounded by the optimal value of the DRO problem. The method is validated via a simple numerical study, which demonstrates improved performance under severely heavy-tailed noise.
\end{abstract}


\section{Introduction}

    System identification seeks to construct mathematical models of dynamical systems from measured input-output (IO) data \cite{LL:99}. For linear time-invariant systems, the data may be analyzed in either the time-domain or the frequency-domain, and in the latter context, frequency-domain identification methods essentially seek to estimate the frequency response function (FRF) of the system  \cite{schoukens_nonparametric_2018}. Classical approaches to system identification include prediction-error methods \cite{PEMs}, which encompass maximum likelihood estimation for appropriate noise models and loss functions, and subspace identification methods, such as MOESP and N4SID  \cite{Verhaegen_Verdult_2007,PVO-BDM:96}. Under suitable regularity conditions, prediction-error methods achieve convergence and asymptotic normality, resulting in large-sample covariance approximations \cite{LL:99,PEMs}. 

    A challenging problem is that of providing finite-sample (i.e., non-asymptotic) guarantees on the quality of the estimated model in the presence of stochastic measurement noise. Early research in finite-sample analysis bounded the difference in expected identification error between the estimated parameters and the optimal parameters \cite{campiFiniteSampleProperties2002}. More recent work has used concentration inequalities and self-normalized martingale techniques to derive finite-sample estimation bounds \cite{matniTutorialConcentrationBounds2019a,tsiamis_statistical_2023,gonzalezFiniteSampleMIMO2025,ziemannTutorialNonAsymptoticTheory2023,honarpishehFrequencyResponseIdentification2026}. While Gaussian, sub-Gaussian, or bounded noise assumptions are common in these analyses \cite{tsiamis_statistical_2023,tsiamisFiniteSampleFrequency2024b, honarpishehFrequencyResponseIdentification2026}, finite-sample guarantees for heavy-tailed noise have also been established under a finite fourth-moment assumption \cite{kanakeriOutlierRobustLinearSystem2025}. 
    
    In contrast with ARX or state-space identification, direct estimation of the FRF has several distinct advantages, including strong connections to both engineering practice \cite{schoukens_nonparametric_2018} and standard control-theoretic uncertainty models \cite{KZ-JCD:98}. Challenges arise in this setting when using non-steady-state data: ignoring the transient contributions will bias the estimated FRF, while waiting for transients to decay may require prohibitively long experiments, particularly for systems with slow dynamics \cite{schoukens_nonparametric_2018}. Conventional approaches accounting for transient and leakage effects are the so-called local polynomial and local rational methods  \cite{schoukens_nonparametric_2018,pintelon2012system,pintelonEstimationNonparametricNoise2010}, which pre-process the data. 

    Leakage-free methods for FRF estimation using non-steady-state data have recently been proposed based on Willems' Fundamental Lemma (WFL), using either time-domain data \cite{ossarehFormulaEstimatingFrequency2023,markovskyFinitedataNonparametricFrequency2024,pedariErrorBoundsDirect2026} or frequency-domain data \cite{Meijer_non-SS_FD_WFL,meijer2026frequencydomainwillemslemmadatadriven}. The methods require no data pre-processing as they implicitly separate the transient and steady-state portions of the response, and they achieve exact FRF recovery in the case of noise-free data. Approaches for the case of noisy output measurements vary: \cite{Meijer_non-SS_FD_WFL,markovskyFinitedataNonparametricFrequency2024,pedariErrorBoundsDirect2026} use a heuristic singular-value decomposition (SVD) truncation to approximate the noise-free data matrix. However, large measurement perturbations --- including outliers under heavy-tailed noise --- can distort singular values and associated directions, and may cause SVD rank truncation to retain directions it should discard, and discard directions it should retain. No out-of-sample guarantees are provided in \cite{Meijer_non-SS_FD_WFL,markovskyFinitedataNonparametricFrequency2024}, while \cite{pedariErrorBoundsDirect2026} provides an error bound under the assumption of bounded measurement noise. Alternatively \cite{ossarehFormulaEstimatingFrequency2023} treats noise through a least squares-inspired projection heuristic, but again no performance guarantees are provided. 


    To summarize, presently no approach to FRF estimation can leverage noisy non-steady-state data without pre-processing while also providing theoretical estimation guarantees on out-of-sample performance with unknown noise distributions. Distributionally robust optimization (DRO) provides a potential approach to address these issues, as it accounts for uncertainty in the unknown noise distribution by minimizing the worst-case expected loss over a set of probability distributions \cite{kuhn2024wassersteindistributionallyrobustoptimization}, and automatically provides finite-sample probabilistic guarantees. DRO has found various applications in control, including robust control design \cite{BVP-DK-PJG-MM:16}, predictive control \cite{JC-JL-FD:22}, and model reduction \cite{liuDistributionallyRobustModel2025}. Applications of DRO to system identification have appeared in \cite{park_robust_2025} for time-domain state-space models, and in \cite{LI2025475} for ARX models, but neither approach provides bounds on the identified FRF error.
    
\medskip


\emph{Contributions:} To seek robust and leakage-free FRF estimates in the presence of unknown measurement noise distributions, this paper develops a Wasserstein DRO framework for frequency-domain system identification from noisy non-steady-state IO data. Using repeated experiments with noisy measurements, we define an empirical probability law for the noisy output spectrum data, and seek robustness against the unknown noise-generating probability law by considering all laws in a weighted Wasserstein ball around the empirical law. Our Wasserstein metric calibration allows for the setting of weights that encode different levels of confidence across experiments, frequency bins (cf. \cite{honarpishehFrequencyResponseIdentification2026}), and output channels. 

We build in particular on the frequency-domain approach of  \cite{Meijer_non-SS_FD_WFL}, and formulate DRO FRF estimation as minimization of a worst-case frequency-domain residual over the weighted ball of distributions. The proposed FRF estimator is then derived by minimizing a convex upper bound on the DRO problem. The approach requires no pre-processing of the non-steady-state data and provides finite data-length estimation error bounds for unknown noise distributions. A simple case study is presented which illustrates the conservatism of the method under light-tailed noise, and advantages of the method in the case of heavy-tailed noise.

\smallskip

\emph{Organization:} Section \ref{Sec:Review} reviews the approach of \cite{Meijer_non-SS_FD_WFL} for the recovery of FRF values based on noise-free non-steady-state frequency-domain data. Section \ref{Sec:UncertaintyModelling} introduces measurement noise into this framework and casts the resulting uncertainty in the framework of Wasserstein ambiguity sets. Section \ref{Sec:Main} introduces our DRO-based FRF estimator and establishes theoretical estimation guarantees. Section \ref{Sec:CaseStudy} presents a numerical case study evaluating our DRO approach, and Section \ref{Sec:Conclusions} concludes the paper.

\smallskip

\emph{Notation:} For $a,b \in \mathbb{Z}$ we let $\mathbb{Z}_{\geq a} = \{a,a+1,\ldots\}$, with $\mathbb{Z}_{[a,b]} = \{a,a+1,\ldots,b\}$. For an index set $\mathcal{I} \subset \mathbb{Z}$, $x = \{x_k\}_{k\in\mathcal{I}}$ denotes the column vector/matrix which stacks the elements $x_k$. The symbol $\otimes$ denotes the Kronecker product of matrices, while $\mathbb{S}_{++}^{n}$ denotes the set of $n \times n$ positive definite matrices The set of eigenvalues of a square matrix $A$ is $\mathrm{eig}(A)$.


\section{Identification with Non-Steady-State Frequency-Domain Data}
\label{Sec:Review}

Consider the discrete-time LTI system
\begin{equation}\label{eqn:lti-system}
    \Sigma: \qquad
    \begin{aligned}
        x_{k+1} &= Ax_k + Bu_k,\\
        y_k &= Cx_k + Du_k,
    \end{aligned}
\end{equation}
where $x_k\in\R^{n_x}$ with initial condition $x_0$, $u_k\in\R^{n_u}$, and $y_k\in\R^{n_y}$ are the state, input, and output, respectively. The matrices $(A,B,C,D)$ are unknown, but $(A,B)$ is assumed to be controllable. The observability index $\ell_\Sigma$ of \eqref{eqn:lti-system} is the smallest integer $\ell\in\Z_{\ge 1}$ such that
    \[
    \operatorname{rank}{\cal O}_\ell(A,C)
    =
    \operatorname{rank}{\cal O}_{n_x}(A,C),
    \]
    where ${\cal O}_\ell(A,C) = \mathrm{col}(C,CA,\ldots,CA^{\ell-1})$. For any $z\in\C\setminus\eig( A)$, let
\begin{equation}
    \label{eqn:transfer-function}
    H(z):=C(zI-A)^{-1}B+D
\end{equation}
denote the transfer function of \eqref{eqn:lti-system}. The broad objective will be to estimate \eqref{eqn:transfer-function} on a grid of frequencies using measured IO data. The remainder of this section follows the formulation derived in \cite{Meijer_non-SS_FD_WFL} for determining $H(z_m)$ from non-steady-state IO frequency-domain data under noise-free conditions.


\subsection{Non-Steady-State Frequency Domain Data}

Let $s := \{s_n\}_{n\in\Z_{[0,2M-1]}}$ be a $\R^{n_{s}}$-valued sequence of length-$2M$ in the time-domain with $M \geq 1$. We define an associated set of frequencies on the uniform grid
\begin{equation}
\label{eqn:freq-grid}
\hat\omega_m = \frac{\pi m}{M} \in [0,\pi),
\qquad 
m\in{\cal M},
\end{equation}
where ${\cal M}:=\{0,\dots,M-1\}$. The DFT $\hat{S} = (\hat{S}_0,\ldots,\hat{S}_{M-1})$ of $s$ is defined component-wise by
\[
\hat S_m
:=
S(\hat\omega_m)
=
\sum_{n=0}^{2M-1}s_ne^{-j\hat\omega_m n}
\in\C^{n_s},
\qquad
m\in{\cal M},
\]
and since $s$ is real-valued, it holds that $\hat S_{-m} =\hat S_m^*$ for all $m\in{\cal M}$, where $\hat S_m^*$ denotes the element-wise complex conjugate of $\hat S_m$. Hence, the negative-frequency coefficients corresponding to $m=-1,\dots,-(M-1)$ are represented implicitly through conjugate symmetry, and $\hat S_0\in\R^{n_s}$. 

Let $u := \{u_n\}_{n\in\Z_{[0,2M-1]}}$ and $y := \{y_n\}_{n\in\Z_{[0,2M-1]}}$ be sequences with corresponding DFTs $\hat{U}$ and $\hat{Y}$. The following definition captures valid input-output spectra.

\begin{definition}[Input-Output Spectrum \cite{Meijer_non-SS_FD_WFL}]
\label{def:IO-spectrum}
    A pair of length-$M$ spectral sequences $(\hat U,\hat Y)$, with $\hat U_m\in\C^{n_u}$ and $\hat Y_m\in\C^{n_y}$, is an IO spectrum of \eqref{eqn:lti-system} if there exists a state sequence $\{x_n\}_{n\in\Z_{[0,2M]}}$ such that its spectrum $\hat X$ satisfies
    \begin{equation}
    \label{eqn:non-SS_IO_spectrum}
    \begin{cases}
        e^{j\hat\omega_m}\hat X_m& = A\hat X_m + B\hat U_m+e^{j\hat\omega_m}(x_0-x_{2M}),
        \\
        \hat Y_m &= C\hat X_m + D\hat U_m
    \end{cases}
    \end{equation}
    for all $m\in{\cal M}$. We call the triple $(\hat U,\hat X,\hat Y)$ an input-state-output spectrum and $(\hat U,\hat X)$ an input-state spectrum.
\end{definition}

    The boundary term $e^{j\hat\omega_m}(x_0-x_{2M})$ in \eqref{eqn:non-SS_IO_spectrum} of Definition \ref{def:IO-spectrum} accounts for the fact that a finite sequence of non-steady-state data is used to compute the spectra \cite{Meijer_non-SS_FD_WFL}. It appears as a consequence of using the DFT and accounts for the mismatch between the initial and terminal states of the measurement window. If $z_m:=e^{j\hat\omega_m}\notin \eig(A)$, then eliminating $\hat{X}_{m}$ from \eqref{eqn:non-SS_IO_spectrum} yields
\[
\hat Y_m=H(z_m)\hat U_m + T(z_m), \quad m \in \mathcal{M},
\]
where $T(z_m)$ denotes the transient
\[
T(z_m):=C(z_m I-A)^{-1}z_m(x_0-x_{2M}).
\]
Since our measured frequency-domain data only consists of pairs of sequences $(\hat U,\hat Y)$, the initial and final states $x_0$ and $x_{2M}$ are unknown, and thus $T(z_m)$ is generally non-zero.


\subsection{Multiple Datasets and Collective Excitation}

In what follows, the input signal spectrum $\hat{U}$ will be required to be sufficiently exciting. For multiple-input multiple-output (MIMO) systems, multiple input signals are required for exciting every input direction \cite{bazanellaClosedLoopIdentificationMIMO2010}. To handle this case, we consider a collection of $E\in\Z_{\ge n_u}$ finite-window IO spectra
\begin{equation}\label{Eq:FD-Data}
\{(\hat U^{(e)},\hat Y^{(e)})\}_{e\in{\cal E}},
\qquad
{\cal E}:=\{1,\dots,E\},
\end{equation}
each of which may be obtained through, e.g., a multi-sine experiment. Importantly, each dataset may have a different boundary-state mismatch and therefore a different transient $T^{(e)}(z_m)$.

\begin{definition}[Data Matrix]
For a collection of sequences $\hat S^{({\cal E})} =\{\hat S^{(e)}\}_{e\in\cal E}$ with $\hat S_m^{(e)}\in\C^{n_s}$ and $L\in\Z_{>\ell_\Sigma}$ let
\[
\begin{aligned}
W_L(z) &:= 
\begin{bmatrix}
    1 & z & \cdots & z^{L-1}
\end{bmatrix}^{\top},
\qquad
z\in\C,\\
    \hat S_m^{({\cal E})}
    &:=
    \begin{bmatrix}
        \hat S_m^{(1)}
        &
        \cdots
        &
        \hat S_m^{(E)}
    \end{bmatrix},\\
{\cal F}_L(\hat S_0)
&:=
\Big[
    W_L(z_0)
    \otimes
    \hat S_0^{({\cal E})}
\Big]\\
{\cal F}_L(\hat S_{[1,M-1]})
&:=
\Big[
    W_L(z_1)
    \otimes
    \hat S_1^{({\cal E})}
    \quad
    \\
    &\cdots
    \quad
    W_L(z_{M-1})
    \otimes
    \hat S_{M-1}^{({\cal E})}
\Big],
\end{aligned}
\]
and
\begin{equation*}
    \label{eqn:Psi-definition}
    \Psi_L(\hat S^{(\mathcal{E})}) 
    :=
    \begin{bmatrix}
        {\cal F}_L(\hat S_0)
        &
        {\cal F}_L(\hat S_{[1,M-1]})
        &
        {\cal F}^*_L(\hat S_{[1,M-1]})
    \end{bmatrix},
\end{equation*}

where $(\cdot)^*$ denotes the element-wise complex conjugation.
\end{definition}

The matrix $\Psi_{L}^{E}(\hat{S})$ aggregates data across all $E$ datasets and all frequency bins.

\begin{definition}[Collective Persistence Excitation]
\label{def:CPE}
Consider a collection of $E$ complex-valued frequency-domain data sequences $\hat S^{({\cal E})}$. The collection is said to be collectively persistently exciting (CPE) of order $L$ if the block matrix $\Psi_L(\hat S^{(\cal{E})}) $ has full row rank (i.e., rank $n_sL$).
\end{definition}

Necessarily, collective persistent excitation of order $L$ requires $E(2M-1)\ge n_sL$.

    
\subsection{Noise-Free Recovery of the FRF}

To capture the effects of the experiment-dependent finite-window transients $T^{(e)}(z_m)$, define the artificial spectral inputs
\begin{equation}
    \hat\Omega_m^{(e)}:=e_e z_m\in\C^E,
\end{equation}
where $e_e$ is the $e$-th canonical vector in $\R^E$, and define the augmented inputs
\[
\hat V_m^{(e)}:=
\col(\hat U_m^{(e)}, \hat\Omega_m^{(e)})
\in\C^{n_u+E}.
\]
The following corollary is a consequence of Theorem 2 from \cite{Meijer_non-SS_FD_WFL} and provides a way to exactly compute $H(z_m)$ in the absence of measurement noise. As notation, for $z_m\notin\eig A$, define
\begin{equation}\label{Eq:calA}
{\cal A}_L(z_m)
:=
\begin{bmatrix}
    0 & \Psi_L(\hat U^{(\cal{E})})
    \\
    0 & \Psi_L(\hat \Omega^{(\cal{E})})
    \\
    -W_L(z_m)\otimes I_{n_y} & \Psi_L(\hat Y^{(\cal{E})})
\end{bmatrix}.
\end{equation}

\smallskip
   
\begin{assumption}
    \label{ass:excited-input-rank}  
    The input evaluation spectra matrix is full rank for all excited frequencies, i.e.,
    \[
    \mathrm{rank}(\hat U_m^{({\cal E})})
        =
            n_u,
    \qquad
    m\in{\cal M}_{\rm ex},
    \]
    where ${\cal M}_{\rm ex}\subseteq {\cal M}$ is a nonempty subset of the frequency grid.
\end{assumption}
\begin{corollary}[Recovery of $H(z_m)$ for MIMO systems]
\label{cor:noise-free-FRF-recovery}
    Consider a collection of IO spectra $\{(\hat U^{(e)}, \hat Y^{(e)})\}_{e\in{\cal E}}$ for the controllable system \eqref{eqn:lti-system}. Suppose that the augmented input collection $\hat V^{({\cal E})}$ is CPE of order $L+n_x$. Then for any $m\in{\cal M}_{\rm ex}$ such that $z_m\notin\eig(A)$, the system of linear equations
\begin{equation}
    \label{eqn:mimo-frf-matrix}
    {\cal A}_L(z_m)
    \begin{bmatrix}
    Y_{m}^{\rm ss}
    \\
    G_m
\end{bmatrix}
=
\begin{bmatrix}
    W_L(z_m)\otimes \hat U_m^{({\cal E})}
    \\
    0
    \\
    0
\end{bmatrix}
\end{equation}
in the unknowns $Y_{m}^{\rm ss}\in\C^{n_y\times E}$ and $G_m \in \C^{E(2M-1)\times E}$ is feasible. The component $Y_{m}^{\rm ss}$ of the solution is unique and satisfies
\[
Y_{m}^{\rm ss}
    =
        H(z_m) \hat U_m^{({\cal E})},
\]
Finally, under Assumption~\ref{ass:excited-input-rank}, the FRF is recovered as 
\[
H(z_m)
    =
        Y_{m}^{\rm ss}  (\hat U_m^{({\cal E})})^\dagger
\in \C^{n_y \times n_u}.
\]
\end{corollary}


\section{Setup for Robust FRF Estimation: Uncertainty Modelling and Wasserstein Uncertainty Sets}
\label{Sec:UncertaintyModelling}

In the noise-free setting, \eqref{eqn:mimo-frf-matrix} recovers $H(z_m)$ directly from frequency-domain data \eqref{Eq:FD-Data}, irrespective of the initial condition, and without explicitly identifying the system matrices in \eqref{eqn:lti-system}. We now consider the case where (i) the output measurement data $\hat{Y}$ is corrupted by noise from an unknown probability distribution, and (ii) the initial condition $x_0$ is modelled as a random variable. The main impact of (i) is that the equation \eqref{eqn:mimo-frf-matrix} no longer holds exactly. We will seek to account for this uncertainty in a robust fashion, and generalize the estimation mechanism of Corollary \ref{cor:noise-free-FRF-recovery} as a distributionally robust optimization problem. The variation in the initial condition will play a more minor role in technical results of Section \ref{Sec:FRFError}.


\subsection{Modelling of Noisy Data}

Suppose that the $E$ experiments leading to the data \eqref{Eq:FD-Data} are repeated $Q\in\Z_{\ge 1}$ times, each experiment beginning from a random initial condition $x_0^{(e,q)} \in \R^{n_x}$. We obtain the IO data
\[
\{(\hat U^{(e)},\hat Y^{(e,q)})\}_{e\in{\cal E},q \in \{1,\ldots,Q\}},
\]
where $\hat Y^{(e,q)} = Y_{\mathrm{sys}}^{(e,q)} + N^{(e,q)}$ is the measured output, consisting of the "noise-free" output spectrum $Y_{\mathrm{sys},m}^{(e,q)}$ and the corrupting additive measurement noise $N^{(e,q)}_m$ at each frequency index $m\in{\cal M}$. Since the initial state is a random variable, the boundary mismatches of the state vary across each repetition and experiment and are denoted as
\[
\Delta x^{(e,q)}
:=
x_0^{(e,q)} - x_{2M}^{(e,q)}.
\]
Each repetition $q$ corresponds to one realization of the complete collection of $E$ measured output spectra. Note that despite the same input $\hat{U}^{(e)}$ being used across the $Q$ repetitions, the noise-free response $Y_{\rm sys}^{(e,q)}$ may vary with $q$ due to changes in initialization of the state.

\begin{assumption}[Experiments]
    \label{ass:repeated-experiment-model}
    The following conditions hold.
    \renewcommand{\labelenumi}{(\roman{enumi})}
    \begin{enumerate}
        \item For every $(e,q)$, there exists a state spectrum $\hat X^{(e,q)}$ such that $(\hat U^{(e)},\hat X^{(e,q)},Y_{\mathrm{sys}}^{(e,q)})$ is an input-state-output spectrum in the sense of Definition \ref{def:IO-spectrum}.
        \item The random variable $x_0^{(e,q)}$ has finite first moment.
    
        \item For all $m\in{\cal M}$, the additive error $N^{(e,q)}_m$ is zero-mean and has finite first moment.
    
        \item The collection
        \[
        \{(x_0^{(e,q)}, N_m^{(e,q)})\}
        \]
        are independent and identically distributed across $q$. 
    \end{enumerate}
\end{assumption}

No independence is imposed among the coordinates within a repetition; hence, the joint output uncertainty may exhibit dependence across experiments, frequency bins, output channels, and real and imaginary parts of spectral coordinates. 

We next seek to organize the noise-corrupted data into vectors suitable for the application of DRO theory. For every $e\in{\cal E}$ and $q\in\{1,\dots,Q\}$, define the $m = 0$ spectral blocks
\[
\hat\eta_{e,0}^{(q)}
:=
\hat Y_0^{(e,q)}
\in\R^{n_y},
\]
and for $m \in \{1,\dots,M-1\}$, define
\[
\hat\eta_{e,m}^{(q)}
:=
\begin{bmatrix}
    \Re(\hat Y_m^{(e,q)})
    \\
    \Im(\hat Y_m^{(e,q)})
\end{bmatrix}
\in\R^{2n_y},
\]
where $\Re(\cdot)$ and $\Im(\cdot)$ denote the real and imaginary components, respectively. For compactness, denote the index set
\[
{ \cal I}
:=
\{(e,m):e\in{\cal E}, m\in{\cal M}\},
\]
and define the indexed dimensions
\[
k_m :=
\begin{cases}
    n_y, & m=0,
    \\
    2n_y, & m=1,\dots,M-1.
\end{cases}
\]
Then for each repetition $q\in\{1,\dots,Q\}$, we stack the noise affected data into the vector 
\[
\hat\xi^{(q)}:=\col\big(\{\hat\eta_{e,m}^{(q)}\}_{(e,m)\in{\cal I}}\big),
\qquad
\hat\eta_{e,m}^{(q)}\in\R^{k_m}.
\]
The vectors $\hat{\xi}^{(q)}$ are realizations of a random variable
\begin{equation}\label{Eq:Defxi}
\xi = \col\big(\{\eta_{e,m}\}_{(e,m)\in{\cal I}}\big), \quad \eta_{e,m}\in\R^{k_m},
\end{equation}
which we suppose satisfies $\xi \in \Xi$ where $\Xi\subseteq\R^{n_\xi}$ is a nonempty closed support set with $ n_\xi:=En_y(2M-1)$. The probability distribution $\mathbb{P}^{\star}$ of $\xi$ is unknown, and will be approximated from a subset of the $Q$ experiments. However, a direct consequence of Assumption \ref{ass:repeated-experiment-model} is that every empirical sample $\hat\xi^{(1)},\dots,\hat\xi^{(Q)}$ is an independent and identically distributed realization of $\Prob^\star$.


\subsection{Weighted Wasserstein Uncertainty Modelling}

While the true probability distribution $\mathbb{P}^{\star}$ of $\xi$ is unknown, an empirical distribution can be estimated through repeated collection of IO data. We can then consider the set of all distributions within a so-called Wasserstein ball of radius $\varepsilon$ around the empirical distribution. The radius of this ball can be calibrated such that $\Prob^\star$ is contained in this set with high probability, and we will thus model $\Prob^{\star}$ as an unknown element belonging to this fixed ball of probability distributions.

    
    We first partition the repeated experiments $\{1,\ldots,Q\}$ into two disjoint subsets $\mathcal{Q}_{\rm cal} \cup \mathcal{Q}_{\rm id} = \{1,\ldots,Q\}$, with $\mathcal{Q}_{\rm cal}$ a calibration set containing the first $Q_{\mathrm{cal}}$ repetitions and $\mathcal{Q}_{\rm id}$ an identification set containing the remaining $Q_{\id}$ repetitions. The set $\mathcal{Q}_{\rm cal}$ allows for calibration of the frequency-dependent weights in our Wasserstein metric, and for now can be left aside. The identification set ${\cal Q}_{\rm id} := \{Q_{\rm cal}+1,\dots,Q\}$ is used to construct the empirical distribution $\hat\Prob$ 
\[
\hat\Prob
:=
\frac{1}{Q_{\rm id}}\sum_{q\in{\cal Q}_{\rm id}}
\delta_{\hat\xi^{(q)}},
\]
where for any measurable set $A \subset \Xi$ 
\[
\delta_{\hat\xi^{(q)}}(A)
=
\begin{cases}
    1, \qquad \hat\xi^{(q)}\in A,
    \\
    0, \qquad \text{otherwise}
\end{cases}
\]
denotes the Dirac probability measure concentrated at the sample $\hat\xi^{(q)}$. The empirical law $\hat\Prob$ captures the distribution of repeated experiments, but will generally differ from the true out-of-sample law $\Prob^{\star}$ when only finitely many experiments are available. To account for this finite-data uncertainty, we consider probability laws contained in a frequency-weighted Wasserstein ball centred at $\hat\Prob$.

\begin{definition}[Weighted Wasserstein Ambiguity Set]
    \label{def:wasserstein-ambiguity-set}
    For every $(e,m)\in{\cal I}$, let $W_{e,m}\in\Sym_{++}^{k_m}$ be a symmetric positive-definite weighting matrix and set
\[
W:=\blkdiag_{(e,m)\in{\cal I}}(W_{e,m})
\in\Sym^{n_\xi\times n_\xi}_{++}.
\]
For $\xi$ as given in \eqref{Eq:Defxi}, the weighted Wasserstein ground metric on $\Xi$ is
    \begin{equation}\label{eqn:GroundMetric}
        \norm{\xi-\xi'}_{W}
        :=
        \sum_{e\in{\cal E}}
        \sum_{m\in{\cal M}}
        \norm{W_{e,m}(\eta_{e,m}-\eta_{e,m}')}^{}_2
        ,
    \end{equation}
where $\xi,\xi'\in\Xi$, and $\norm{\cdot}_2$ is the Euclidean norm.

Let ${\cal P}_{}(\Xi)$ denote the set of probability laws on $\Xi$ with finite first moment under $\norm{\cdot}_{W}$. Then for any two probability laws $\Prob_1,\Prob_2 \in \mathcal{P}(\Xi)$, the Wasserstein distance induced by $\|\cdot\|_{W}$ is
\begin{equation}\label{Eq:Wd}
d_{}(\Prob_1,\Prob_2)
:=
\inf_{\Pi\in{\cal C}(\Prob_1,\Prob_2)}
\int_{\Xi \times \Xi}
\norm{\xi- \xi'}_{W}^{}
\,
\Pi(d\xi,d\xi'),
\end{equation}
where ${\cal C}(\Prob_1,\Prob_2)$ denotes the set of couplings of $\Prob_1$ and $\Prob_2$, and for $\varepsilon_{}>0$, the weighted Wasserstein ambiguity set is 
\begin{equation}
\label{eqn:wass-ball}
    {\cal B}_{\varepsilon_{}}(\hat\Prob)
    :=
    \left\{
          \Prob\in{\cal P}_{}(\Xi)
          \mid
          d_{}(\Prob,\hat\Prob)
          \le
          \varepsilon_{}
    \right\}.
\end{equation}
\end{definition}

    See \cite{kuhn2024wassersteindistributionallyrobustoptimization} for a complete description of Wasserstein theory. Each empirical sample contains the complete collection of output spectra and therefore retains any joint dependencies across datasets and frequencies, as permitted by Assumption~\ref{ass:repeated-experiment-model}. The block-diagonal ground metric in \eqref{eqn:GroundMetric} allows for different weightings across experiments and frequencies $(e,m)$. Accordingly, cross-output and real-imaginary dependencies are represented within each block, while cross-frequency and cross-dataset dependencies are not explicitly incorporated into the transport geometry. This structure is adopted to reduce block-weighting-estimation requirements and preserve computational tractability. The weighting matrix $W$ allows us to shape the geometry of the Wasserstein ball, such that the tolerated variability of $\xi$ in the considered ambiguity set is non-uniform and can be calibrated using collected data.


\section{Main Results: Distributionally Robust FRF Estimation}
\label{Sec:Main}

We now seek to develop a distributionally robust counterpart to the FRF estimation mechanism in Corollary \ref{cor:noise-free-FRF-recovery}, wherein the effect of noise from the unknown distribution $\mathbb{P}^{\star}$ is accounted for by robust optimization using the Wasserstein ambiguity set $\mathcal{B}_{\epsilon}(\hat{\mathbb{P}})$ around the empirical distribution $\hat{\mathbb{P}}$.


\subsection{Formulation of DR FRF Estimator}

    We return to the linear equations \eqref{eqn:mimo-frf-matrix} with $\mathcal{A}_{L}(z_m)$ as defined in \eqref{Eq:calA}, but now with output measurements corrupted by noise as described in Section \ref{Sec:UncertaintyModelling}. 

    Fix an evaluation point $m\in{\cal M}_{\rm ex}$ with $z_m\notin\eig(A)$, and for convenience re-label the variables in \eqref{eqn:mimo-frf-matrix} as $\hat\theta_m:=(\hat Y_{m}^{\rm ss},\hat G_m)$. The prescribed input spectra $\hat U^{({\cal E})}$ and transient spectra $\hat\Omega^{({\cal E})}$ are known exactly, so we extract the corresponding rows of  \eqref{eqn:mimo-frf-matrix} as deterministic constraints
    \begin{equation}
    \label{eqn:deterministic-constraints}
    \begin{aligned}
    \Psi_L(\hat U^{(\cal{E})}) \hat G_m &= W_L(z_m)\otimes \hat U_m^{({\cal E})},\\
    \Psi_L(\hat \Omega^{(\cal{E})}) \hat G_m &= 0_{EL\times E}
    \end{aligned}
    \end{equation}
    and let
\begin{equation} 
\label{eqn:psi-estimation-feasible-set}
\begin{aligned}
\Theta_m:=
    \Big\{
        \hat\theta_m = (\hat Y_{m}^{\rm ss},\hat G_m)
        \mid
        \eqref{eqn:deterministic-constraints}\,\,\text{holds}.
    \Big\}
\end{aligned}
\end{equation}
denote the feasible set induced by these constraints. 

With noisy output data, the coefficient matrix $\Psi_{L}(\hat{Y}^{(\mathcal{E})})$ in the third block row of equations in \eqref{eqn:mimo-frf-matrix} will be perturbed by noise, and will generally become full rank; exactly solving this equation no longer guarantees recovery of the true FRF. Define its residual
\begin{equation}\label{Eq:DRResidual}
    r_Y(\xi,\hat\theta_m)
    :=
    \Psi_L(\xi) \hat G_m
    -
    K_m \hat Y_{m}^{\rm ss},
\end{equation}
where for compactness we have set
\[
K_m
    :=
        W_L(z_m)\otimes I_{n_y},
\qquad
\Psi_L(\xi)
    :=
    \Psi_L\big(Y^{(\cal{E})}(\xi)\big).
\]
Note that $\Psi_{L}$ is in fact a linear function of $\xi$. The cost to be minimized will be the Frobenius norm of the residual matrix
\begin{equation}\label{Eq:DRCost}
c(\xi,\hat\theta_m)
:=
\|r_{Y}(\xi,\hat\theta_m)\|_F.
\end{equation}
The worst-case expected value of the random cost over the weighted Wasserstein ball \eqref{eqn:wass-ball} is then
\begin{equation}\label{Eq:JThetaM}
J(\hat\theta_m)
:=
\sup_{\Prob\in{\cal B}_{\varepsilon_{}}(\hat\Prob)}
    \E_\Prob\left[
        c(\xi,\hat\theta_m)
    \right],
\end{equation}
and the corresponding DR estimation problem is to minimize $J(\hat\theta_m)$ subject to the deterministic constraints \eqref{eqn:psi-estimation-feasible-set}, i.e.,

\begin{equation}
\label{eqn:dr-frf}
\begin{aligned}
    \inf_{
       (\hat Y_{m}^{\rm ss},\hat G_m) \in \Theta_{m}}
    &\sup_{\Prob\in{\cal B}_{\varepsilon}(\hat\Prob)}
    \E_\Prob\left[
        \|
            \Psi_L(\xi) \hat G_m
            -
            K_m \hat Y_{m}^{\rm ss}
        \|_F
    \right],
\\
\text{subject to} \quad
    &\Psi_L(\hat U^{(\cal{E})}) \hat G_m = W_L(z_m)\otimes \hat U_m^{({\cal E})},\\
    &\Psi_L(\hat \Omega^{(\cal{E})}) \hat G_m = 0_{EL\times E}.
\end{aligned}
\end{equation}
The maximization in \eqref{Eq:JThetaM} is infinite-dimensional before reformulation. However, the optimal value of $J(\hat\theta_m)$ admits a convex upper bound given by a regularized function of the empirical data \cite[Theorem 5]{kuhn2024wassersteindistributionallyrobustoptimization}. 

To state this upper bound, examining \eqref{Eq:DRResidual}-\eqref{Eq:DRCost}, the first term in the complex-valued residual \eqref{Eq:DRResidual} is bilinear in $\xi$ and $\hat{\theta}_m$, while the second term is linear in $\hat{\theta}_m$. One may therefore construct linear operators $M(\hat{\theta}_m)$ and $b(\hat{\theta}_m)$ such that
\begin{equation}\label{Eq:CasM}
c(\xi,\hat\theta_m) = \|M(\hat\theta_m)\xi - b(\hat\theta_m)\|_2.
\end{equation}
See the Appendix for the explicit construction. Consistent with the partitioning of the matrix $W$ in Definition \ref{def:wasserstein-ambiguity-set}, let $M_{e,k}(\hat\theta_m)$ denote the set of columns of $M(\hat\theta_m)$ associated with $(e,k) \in \mathcal{I}$. We can now give the upper bound.

\begin{theorem}[DR FRF Estimation]
\label{thm:psi-frf-w1}
For every $\hat\theta_m\in\Theta_m$ it holds that $J(\hat\theta_m) \leq \bar{J}(\hat\theta_m)$, where
\begin{subequations}
\begin{align}
\label{eqn:tractable-upper-bound}
\bar{J}(\hat\theta_m) &:=
    \frac{1}{Q_{\rm id}}\sum_{q\in{\cal Q}_{\rm id}}
    \norm{M(\hat\theta_m)\hat\xi^{(q)} - b(\hat\theta_m)}_2
    +
    \varepsilon_{}{\cal L}(\hat\theta_m)\\
\label{eqn:psi-w1-block-induced-norm}
{\cal L}(\hat\theta_m)
    &:=
        \max_{(e,k)\in{\cal I}}
        \norm{M_{e,k}(\hat\theta_m) W_{e,k}^{-1}}_2.
\end{align}
\end{subequations}
Furthermore, if $\Xi=\R^{n_\xi}$, then $J(\hat\theta_m) = \bar{J}(\hat\theta_m)$.
\end{theorem}

The proof of Theorem \ref{thm:psi-frf-w1} is given in the appendix. Based on the upper bound, the proposed distributionally robust estimator is
\begin{equation}
\label{eqn:optimal-frf-estimate}
    \hat\theta_m^\star
    =
    (\hat Y^{\mathrm{ss},\star}_m,\hat G_m^\star)
    \in
    \arg\min_{\theta_m\in\Theta_m}
    \bar{J}(\hat\theta_m).
\end{equation}
The problem \eqref{eqn:optimal-frf-estimate} is convex since $\Theta_m$ is an affine set, each empirical residual in \eqref{eqn:tractable-upper-bound} is affine in $\hat\theta_m$, and ${\cal L}(\hat\theta_m)$ is a maximum of spectral norms of matrices that depend linearly on the component $\hat G_m$ of $\hat{\theta}$. In particular, \eqref{eqn:optimal-frf-estimate} is a semidefinite program (SDP), and admits an exact reformulation as a second-order cone program (SOCP) when $n_y = E = 1$. In the general case $n_y \geq 2$, a more conservative SOCP approximation may be obtained by replacing the regularizer $\mathcal{L}(\hat{\theta}_m)$ with
\[
{\cal L}_F(\hat\theta_m)
    :=
        \max_{(e,k)\in{\cal I}}\norm{M_{e,k}(\hat\theta_m) W_{e,k}^{-1}}_F,
\]
such that the resulting objective is an upper bound on \eqref{eqn:tractable-upper-bound}. 
 
Under the CPE condition of Corollary \ref{cor:noise-free-FRF-recovery}, $\Theta_m$ is a nonempty affine set, and thus there exists $\hat\theta_m^\star$ that minimizes the tractable upper bound \eqref{eqn:tractable-upper-bound}. 
Then under the full rank condition of Assumption \ref{ass:excited-input-rank}, we construct a corresponding FRF estimate as
\begin{equation}\label{Eq:FRFEstimator}
\hat H^\star(z_m)=\hat Y_{m}^{\rm ss,\star} (\hat U_m^{({\cal E})})^\dagger.
\end{equation}

   Note that the weights $W_{e,k}$ used in defining the metric in Definition \ref{def:wasserstein-ambiguity-set} appear in the regularizer $\mathcal{L}(\hat{\theta}_m)$ in Theorem \ref{thm:psi-frf-w1}, and these weights are allowed to vary across experiments $e\in{\cal E}$ and frequencies $k \in \mathcal{M}$. Choosing smaller weights permits larger displacements per unit transport cost (cf. \eqref{Eq:Wd}), and the regularizer discourages solutions that are sensitive to those displacements. The radius $\varepsilon$ determines the trade-off between empirical residual fit and this largest weighted sensitivity. This combination ensures that uncertainty is only penalized when it can actually affect the residual.
   

\subsection{FRF Error Analysis}
\label{Sec:FRFError}

The DR FRF estimator \eqref{eqn:optimal-frf-estimate}--\eqref{Eq:FRFEstimator} minimizes the worst case expected norm of the residual \eqref{Eq:DRResidual}, and then computes an associated estimate $\hat{H}^{\star}(z_m)$ of the FRF. The next result establishes a theoretical bound between the estimate $\hat{H}^{\star}(z_m)$ and the true FRF value $H(z_m)$; the proof can be found in the Appendix.

\begin{proposition}[Bound on Estimation Error]
\label{Prop:BoundEstimationError}
Suppose that Assumption \ref{ass:excited-input-rank}, Assumption \ref{ass:repeated-experiment-model}, and the conditions of Corollary \ref{cor:noise-free-FRF-recovery} hold. Let $\bar{J}^{\star}$ denote the optimal value of \eqref{eqn:optimal-frf-estimate}. If $\Prob^\star \in {\cal B}_{\varepsilon}(\hat\Prob)$, then for each $m\in{{\cal M}_{\rm ex}}$, the FRF estimate $\hat{H}^{\star}(z_m)$ produced by \eqref{eqn:optimal-frf-estimate}--\eqref{Eq:FRFEstimator} satisfies
\begin{equation}
\label{eqn:type-1-frf-error-bound}
\begin{aligned}
    \|\hat H^{\star}(z_m) - H(z_m)\|_F 
    \leq
    \gamma_m \bar{J}^{\star}
    \qquad
\end{aligned}
\end{equation}
for some constant $\gamma_m > 0$.
\end{proposition}

The bound \eqref{eqn:type-1-frf-error-bound} shows that the FRF estimation error is bounded by the optimal value of the problem \eqref{eqn:optimal-frf-estimate}; we emphasize however that \eqref{eqn:type-1-frf-error-bound} is not a consistency result, as $\bar{J}^{\star}$ will generally not tend to zero as $\varepsilon \to 0$.

\begin{remark}[Calibration of $\varepsilon$ Statistical Guarantees]
\label{rmk:statistical-guarantees}
    Proposition \ref{Prop:BoundEstimationError} is stated under the containment hypothesis $\Prob^\star \in {\cal B}_{\varepsilon}(\hat\Prob)$, but probabilistic statements can instead be made. If the unknown law $\Prob^\star$ is light-tailed in the sense that there exist constants $a > 1$ such that $\E_{\Prob^\star}[\exp(\norm{\xi}_{W}^a)]$ is finite, then for $\beta \in (0,1)$, \cite[Theorem~18]{kuhn2024wassersteindistributionallyrobustoptimization} provides a sufficient radius $\varepsilon_{Q_{\rm id}}(\beta)$ such that if $\varepsilon\ge\varepsilon_{Q_{\rm id}}(\beta)$, then $\Prob^\star \in {\cal B}_{\varepsilon}(\hat\Prob)$ with probability at least $1-\beta$. Under these conditions, the bound \eqref{eqn:type-1-frf-error-bound} holds with the same confidence level $1-\beta$. However, the sufficient radius can be conservative \cite[Remark~5]{kuhn2024wassersteindistributionallyrobustoptimization}; this calibration method is not used in the numerical case study below, and instead $\varepsilon$ is tuned empirically.

Note that standard normal Gaussian distributions satisfy this exponential-moment condition. While Student-$t$ laws have finite first absolute moments, their variance may be infinite. Cauchy noise additionally lacks a finite first moment, meaning Assumption~\ref{ass:repeated-experiment-model} no longer holds. We therefore include these distributions as empirical stress tests, with Cauchy explicitly outside of the DR uncertainty modelling framework. \hfill \oprocend

\end{remark}


\section{Numerical Case Study}
\label{Sec:CaseStudy}

    This section implements our FRF estimator on the benchmark system described in \cite{Meijer_non-SS_FD_WFL,McKelvey-1996} and compares its performance under various frequency-dependent measurement noise models. The case study will address the following questions: 
    \begin{itemize}
        \item How does the DRO FRF estimator compare to an averaging-based baseline estimator?
        \item How does the comparison change as measurement noise trends from light-tailed to heavy-tailed?
    \end{itemize}

\subsection{Plant and Experimental Data}
We implement the fourth-order discrete-time SISO benchmark plant used in  \cite{Meijer_non-SS_FD_WFL} with transfer function
\[
H(z)=
\frac{9.626 z^4 + 4.095 z^3 - 9.718 z^2 + 2.6 z + 8.618}
{10 z^4 - 3.306 z^3 - 5.025 z^2 - 2.347 z + 7.925}.
\]
which displays two strong resonances, one at $\omega \approx 0.15\pi$ and one at $\omega \approx 0.76\pi$. We perform $Q=7$ repetitions of $E = 1$ experiments over $2M=56$ time samples, each simulated from $x_0=0$ each under the same input in every repetition, with $M=28$ retained DFT bins, using the grid described in \eqref{eqn:freq-grid}. Consequently, we solve an exact SOCP reformulation of \eqref{eqn:tractable-upper-bound}, retaining its original spectral-norm regularizer. Adopting the input excitation procedure from \cite{Meijer_non-SS_FD_WFL}, we excite odd frequencies, (i.e. ${\cal M}_{\rm ex}=\{1,3,\dots,M-1\}$) with a prescribed excited input of $\hat U_m=1$ for all $m\in{\cal M}_{\rm ex}$, and $\hat U_m=0$ otherwise. We use $L = 5$ for the depth of the data matrices. Note that all $m\in{\cal M}$ retained frequencies are used to construct $\Psi_L(\hat U^{(\cal{E})})$ and $\Psi_L(\hat Y^{(\cal{E})})$, while FRF estimates will be only be evaluated at excited frequencies $m\in\mathcal{M}_{\rm ex}$.


\subsection{Noise Models}

    Following the noise modelling in Section \ref{Sec:UncertaintyModelling}, noise is applied directly to the output DFTs, with a real DC coefficient and conjugate-symmetric negative-frequency coefficients. We consider noise distributions with progressively heavier tails, chosen to examine two distinct challenges: frequency-dependent measurement quality and robustness to large perturbations. The examined noise laws include Gaussian, Student-$t$, and Cauchy distributions. 

\smallskip


\subsubsection{Frequency-Dependent Noise Scaling}

To model increased noise amplitudes being present at higher frequencies, we apply the sigmoid logistic envelope
    \[
    \sigma_{m}
        :=
            0.03
            +
            \frac
                {0.62}
                {1+ \exp\left[
                    -35\left(
                        \frac{\hat\omega_m}{\pi}
                        -0.36
                    \right)
                \right]}.
    \]
This envelope smoothly increases a multiplier that scales the noise from approximately $0.03$ to $0.65$, with its midpoint occurring at $0.36\pi$, where $\sigma_m=0.34$.

\subsubsection{Frequency-Dependent Noise Distributions}
We consider the Student-$t$ spectral noise distribution with $\nu \in \mathbb{R}_{>0} \cup \{\infty\}$ degrees of freedom
\begin{equation}\label{Eq:Student}
T_m^{(q)}(\nu)
:=
Z_m^{(q)}
\sqrt{
            \frac
                {\nu}
                {V_{\nu}^{(q)}}},
\end{equation}
with $Z_0^{(q)}\sim{\cal N}(0,1)$, and for $m \in \{1,\dots,M-1\}$,
\[
Z_m^{(q)}
:=
\frac
    {Z_{\Re,m}^{(q)} + jZ_{\Im,m}^{(q)}}
    {\sqrt 2},
\]
where $Z_{\Re,m}^{(q)}, Z_{\Im,m}^{(q)} \sim \mathcal{N}(0,1)$, and where $V_\nu^{(q)}\sim\chi_\nu^2$ is independent of $Z_m^{(q)}$. Note that \eqref{Eq:Student} introduces dependence across frequency bins within repetition $q$, while the repetitions remain independent, in accordance with Assumption~\ref{ass:repeated-experiment-model}. For all $m\in{\cal M}$, the simulated noise laws are then expressed as
\[
N_m^{(q)}(\nu)
    :=  
        a_\nu\sigma_{m}
        T_m^{(q)}(\nu),
\qquad
a_\nu
    :=
        \frac{\Phi^{-1}(0.75)}{F_{t_\nu}^{-1}(0.75)}.
\]
where  $\Phi$ and $F_{t_\nu}$ denote the cumulative distribution functions of the standard normal and Student-$t$ distributions, respectively, and their inverses are quantile functions. The constant $a_\nu$ normalizes the frequency-weighted heavily-tailed noise $\sigma_{m}T_m^{(q)}(\nu)$ to have identical marginal median absolute deviation to a standard Gaussian real noise coordinate at the same frequency. This fixes the typical noise spread across all distributions, allowing for comparison of tail behaviour without a few extreme realizations distorting the scale. This is in contrast with variance-based normalization, which cannot be done for this study as the second moment of $T_m^{(q)}(\nu)$ is infinite for severely heavy-tailed distributions ($\nu\le 2$).

To investigate the effect of tail heaviness, we vary the degrees of freedom $\nu$ and consider $\nu$ in the set
\[
{\cal V} 
    :=
        \{\infty, 5, 3, 1.8, 1.5, 1.2, 1\}.
\]
This includes Gaussian $(\nu=\infty)$, finite-variance Student-$t$ $(\nu>2)$, infinite-variance Student-$t$ $(1<\nu\le2)$, and Cauchy $(\nu=1)$ distributions. As $\nu$ decreases, the corresponding distribution becomes heavier tailed, meaning that significant perturbations are increasingly common. 


    
\subsection{Implementation of DRO Estimator and Comparison}

\begin{figure}
    \centering
    \includegraphics[width=1\linewidth]{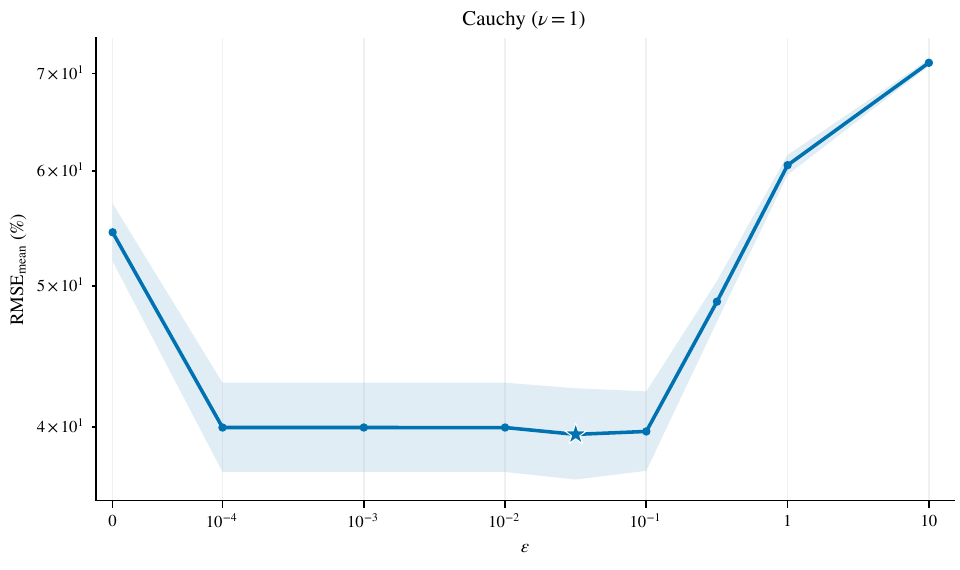}
    \caption{Comparison of mean RMSE percentages for identity-weighted DR estimation across all $\varepsilon\in{\cal E_B}$ under a Cauchy noise distribution $(\nu=1)$. The optimal radius $\varepsilon^\star$ is marked with a star and subsequently used to compare results.}
    \label{fig:epsilon-sweeps}
\end{figure}

The DRO FRF estimator \eqref{eqn:optimal-frf-estimate}--\eqref{Eq:FRFEstimator} is implemented as follows. The weighting matrix in Definition \ref{def:wasserstein-ambiguity-set} is taken for this study as identity $W=I_{n_\xi}$; as such we take $Q_{\rm cal}=0$ and $Q_{\rm id}=Q$. To investigate the impact of the choice of the Wasserstein ball radius $\varepsilon$, for each $\nu\in {\cal V}$, we sweep through the set of radii $\varepsilon \in {\cal E_B}$, where
\[
{\cal E_B}
:=
\{
    0,
    10^{-4},
    10^{-3},
    10^{-2},
    10^{-3/2},
    10^{-1},
    10^{-1/2},
    1,
    10
\},
\]
solve \eqref{eqn:optimal-frf-estimate}, compute \eqref{Eq:FRFEstimator}, and calculate the relative FRF RMSE
\begin{equation}
\label{eqn:FRF-RMSE}
e_r(\varepsilon)
    :=
    \sqrt{
        \frac
            {\sum_{m\in{\cal M}_{\rm ex}}
                \norm{\hat H_r^\star(z_m,\varepsilon) - H(z_m)}_2^2}
            {\sum_{m\in{\cal M}_{\rm ex}}
                \norm{H(z_m)}_2^2}.
    }
\end{equation}
where $r$ indexes the seed used for generating random noise; We repeat this process using $R=96$ distinct seeds, and compute the average RMSE
\begin{equation}
\label{eqn:mean-rmse}
\mathrm{RMSE}_{\rm mean}(\varepsilon) := \frac{100}{R}
        \sum_{r=1}^{R}
        e_{r}(\varepsilon)
\end{equation}
Figure \ref{fig:epsilon-sweeps} plots this as a function of $\varepsilon$ for the case of Cauchy noise. The shaded region shows the pointwise $95\%$ percentile-bootstrap ranges from 5,000 paired whole-trial samples.

A small positive radius reduces error relative to the unregularized solution, while performance deteriorates under excessive regularization. Similar qualitative regularization trade-offs have been observed in DR DeePC \cite{Coulson2019CDC}, although its control objective differs from the identification error considered here. We observe that any selection in the range $[10^{-4},10^{-1}]$ produces comparable performance. Interestingly, this trend was seen across all $\nu\in{\cal V}$, and suggests that the user has flexibility in selecting $\varepsilon$. For concreteness, in our tests we select $\varepsilon^\star$ as the minimizer of this curve. The optimal radius is then fixed independently for each $\nu$.





We compare the DRO FRF estimator against the SVD-based identification framework of \cite[Algorithm 1]{Meijer_non-SS_FD_WFL}, applied to the averaged output data
\begin{equation}
\label{eqn:avg-output-spectrum}
    \bar Y := \frac{1}{Q}\sum_{q=1}^{Q}\hat Y^{(q)}.
\end{equation}
The true order $n_x=4$ is provided, and the method retains $2L+n_x=14$ left singular vectors.

\subsection{Statistical Comparison and Metrics}

We now compare the estimators on an additional $R$ noise seeds that are disjoint from those used in Wasserstein radius selection. For both the DR and averaged-SVD methods, we compute the RMSE via \eqref{eqn:FRF-RMSE} (for the DRO method, $\varepsilon$ is now fixed at $\varepsilon^{\star}$), and report
    \[
    \begin{aligned}
    \mathrm{RMSE}_{{\rm med}}
    &:=
        100 \operatorname{median}_r
        e_{r},\\
    {\mathrm{CVaR}}_{0.9}
        &:=
            100\min_{\tau\in\R}
            \left\{
                \tau +
                \frac{1}{0.1R}
                \sum_{r=1}^{R}
                ( e_{r} - \tau)_+
            \right\},
            \end{aligned}
    \]
    respectively, where $(x)_+:=\max\{x,0\}$, in addition the mean relative RMSE \eqref{eqn:mean-rmse}. These two additional metrics are the median and the   conditional value-at-risk (CVaR),  which computes the mean of the upper $10\%$ of trial errors.

\begin{figure}[ht!]
    \centering
    \includegraphics[width=1\linewidth]{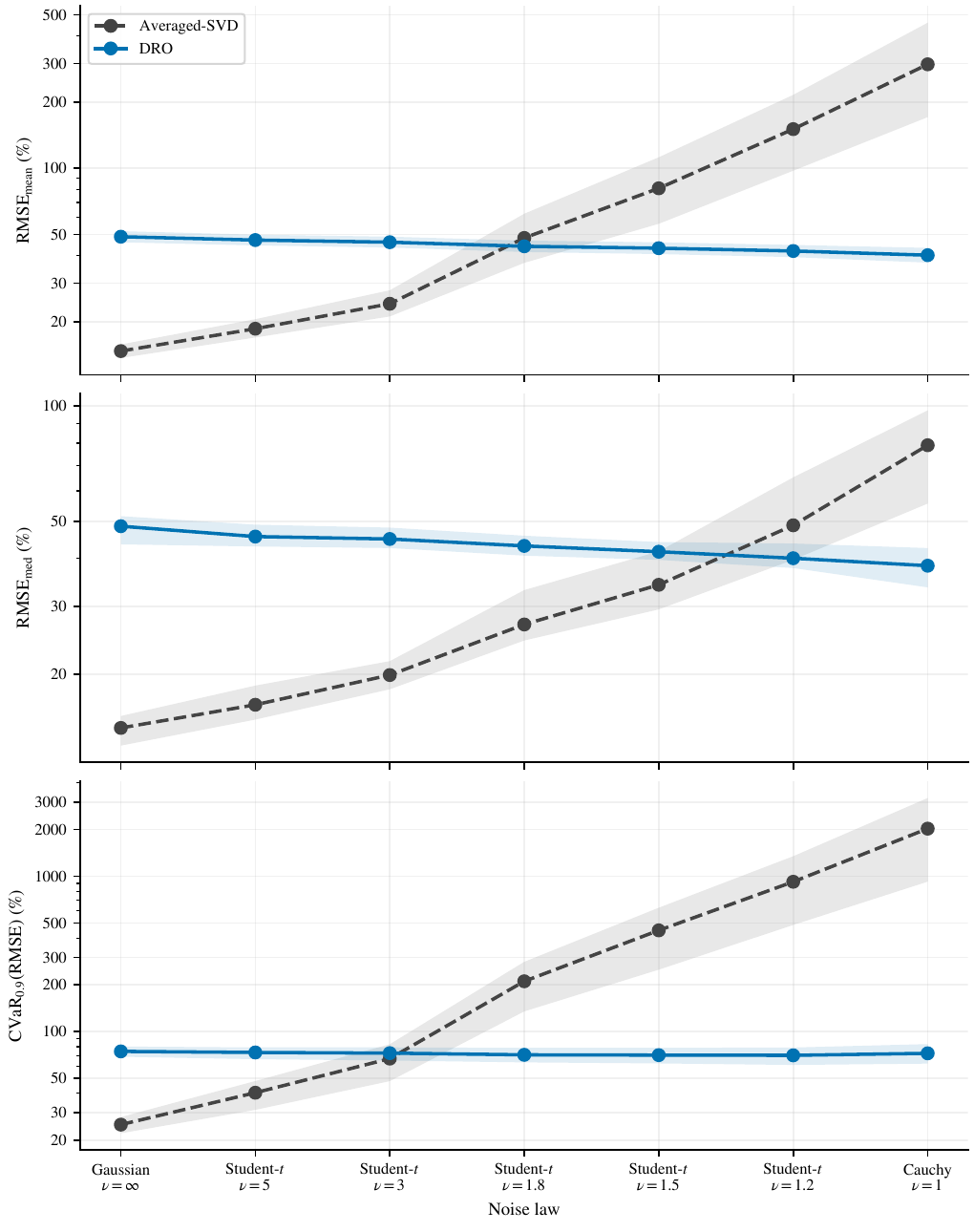}
    \caption{Comparison of mean, median, and $\mathrm{CVaR}_{0.9}$ RMSE percentages for averaged-SVD and DR estimators across all $\nu\in{\cal V}$ for fixed $Q=7$. Note that the optimal Wasserstein radii were selected independently for each $\nu$.}
    \label{fig:metrics}
\end{figure}

Figure~\ref{fig:metrics} plots these three metrics across the different noise distributions $\nu \in \mathcal{V}$, from light-tailed on the left to heavy-tailed on the right. Averaged-SVD has lower mean, median and CVaR point estimates under Gaussian and Student-$t$ noise with $\nu\in\{\infty,5,3\}$. However, as $\nu$ decreases, its observed errors increase substantially, whereas DR mean RMSE decreases from $48.85\%$ to $40.25\%$. Interestingly, the DR CVaR remains relatively stable around $70\%$ across all $\nu\in{\cal V}$. For Cauchy noise, mean RMSE decreases from $297.26\%$ for averaged-SVD to $40.25\%$ for DR. Hence, the strongest benefit of DR is its robustness under severely heavy-tailed noise distributions.

Figure \ref{fig:metrics} displays a crossover between the two mean RMSE errors at $\nu \approx 2$; for larger $\nu$, averaging appears to benefit the SVD-based method, while for smaller $\nu$ it does not. Additionally, while both achieve similar mean RMSE at $\nu\approx 2$, we see a tradeoff; DRO has higher median error yet lower CVaR. This can be explained as follows. Since the clean response and transient are identical across repetitions, averaging can preserve both while reducing measurement noise, ultimately strengthening the baseline. Denoting
\[
\bar N_m 
    :=
        \frac{1}{Q}\sum_{q=1}^{Q}
        N_m^{(q)},
\]
we have, for each fixed frequency $m\in{\cal M}$, independent and identically distributed noise samples across repetitions (Assumption \ref{ass:repeated-experiment-model}). For zero-mean noise from a Student-$t$ distribution with finite variance $(\nu > 2)$,
\begin{equation}
\begin{aligned}
\E[|\bar N_m|^2]
    &=
        \frac1Q
        \E[|N_m^{(q)}|^2] =
        \frac{a_\nu^2\sigma_{m}^2}{Q}
        \frac{\nu}{\nu-2},
\end{aligned}
\end{equation}
for all $q\in{\cal Q}$ \cite{RJ-2013-033}. Since $\bar N_m =\bar Y_m - Y_{{\rm sys},m}$, the averaged output converges to $Y_{{\rm sys},m}$ as $Q\to \infty$, with RMS averaging error
\[
\sqrt{\E[|\bar Y_m - Y_{{\rm sys},m}|^2]}
    =
        \frac
            {a_\nu\sigma_{m}}
            {\sqrt Q}
        \sqrt{\frac{\nu}{\nu-2}}.
\]
For the Gaussian noise $(\nu=\infty)$, the RMS averaging error is exactly $\sigma_m/\sqrt{Q}$ since $a_\infty=1$ and $T_m^{(q)}(\infty)=Z_m^{(q)}$. 

While averaging continues to help mitigate noise for Student-$t$ distributions with infinite variance $(1<\nu\le2)$, the typical averaging error decreases more slowly. Consequently, averaging can remain strongly affected by extreme records when data is scarce. For Cauchy noise $(\nu=1)$, the average of independent, identically distributed real noise coordinates has the same distribution as one coordinate, so averaging does not reduce noise \cite{durrett2019probability}.

Frequency-wise diagnostics discern how error is reduced across the excited spectrum. We calculate the RMS magnitude and phase errors at each excited frequency for both the averaged-SVD and DR methods
\[
|H_{m}|^{\rm RMS}
    :=
        \sqrt{
            \frac1R
            \sum_{r=1}^{R}
            \Big(|\hat H_{r}(z_m)| - |H(z_m)|\Big)^2
        },
\]
\[
\Phi_{m}^{\rm RMS}
    :=
        \sqrt{
            \frac
                {1}{R}
            \sum_{r=1}^R
            \left[
                \operatorname{Arg}
                \{ \hat H_{r}(z_m)  (H(z_m))^* \}
            \right]^2
        }.
\]

\begin{figure}[ht!]
    \centering
    \includegraphics[width=1\linewidth]{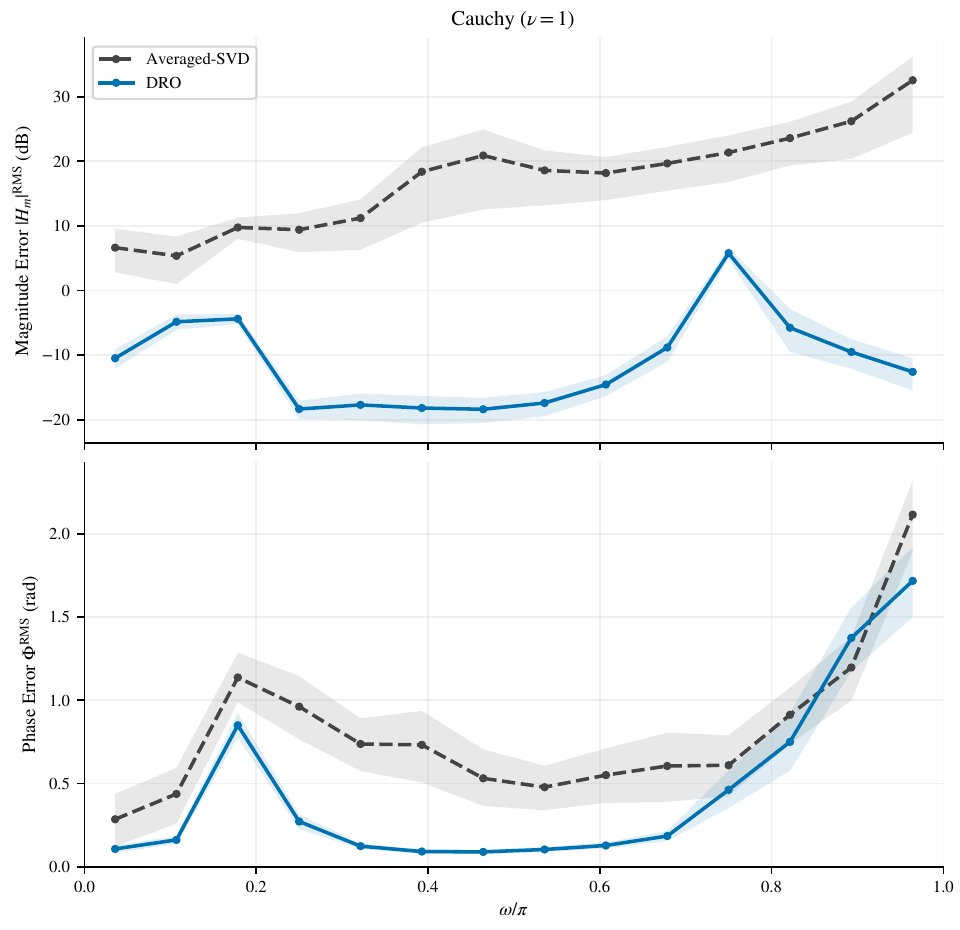}
    \caption{Comparison of RMS FRF and phase errors for averaged-SVD and DR methods under Cauchy noise $(\nu=1)$.}
    \label{fig:FRF-phase-RMS}
\end{figure}

The frequency-wise FRF magnitude errors are plotted in decibels $20\log_{10}( |H_{m}|^{\rm RMS})$ and the phase errors in radians. Under Cauchy noise, Figure~\ref{fig:FRF-phase-RMS} shows lower FRF estimation error than averaged-SVD across all $14$ evaluated frequencies, and lower phase estimation error at $13$ of them; the exception is $\hat\omega_m=25\pi/28$. Note that the DR estimation errors spike around the resonance peak frequencies, reflecting the sensitivity around this region.

\section{Conclusion}
\label{Sec:Conclusions}
This paper developed a DRO framework for frequency-response estimation using noisy, non-steady-state IO data. The resulting convex estimator combines empirical residual fitting with a weighted sensitivity penalty, providing a principled connection between spectral uncertainty and regularization. In a numerical case study comparing to an averaged SVD-based estimator, the DRO-based estimator achieved substantially improved performance across several metrics for heavy-tailed noise, while underperforming the SVD estimator for light-tailed noise. These findings support its potential for limiting large identification errors when repeated measurements contain extreme perturbations.

%

Future work will assess the optimal calibration of weights $W_{e,m}$ such that the frequency-dependent Wasserstein metric obtains superior results to the identity-weighted case. The tradeoff between the number of repetitions used for calibration versus FRF identification will then also be studied. Additionally, the performance under varying initial condition $x_0$, allowed by Assumption~\ref{ass:repeated-experiment-model}, will be assessed.


\section*{References}
\bibliographystyle{IEEEtran}
\begingroup
  \renewcommand{\section}[2]{}
  \bibliography{alias,references}
\endgroup

\appendix


\emph{Construction of $M(\hat\theta_m)$ and $b(\hat\theta_m)$:} Given a complex matrix $R$, define the realification operator
\[
{\frak r}(R)
:=
\begin{bmatrix}
    \Re\vec(R) \\ \Im\vec(R)
\end{bmatrix},
\qquad
\norm{{\frak r}(R)}_2 = \norm{R}_F
\]
Then
\begin{equation}
\begin{aligned}
    c(\xi,\hat\theta_m)
    &=
        \norm{{\frak r}(r_Y(\xi,\hat\theta_m))}_2  \\
    &=
        \norm{
            {\frak r}(\Psi_L(\xi) \hat G_m)
            -
            {\frak r}( K_m \hat Y_{m}^{\rm ss})
        }_2 
\end{aligned}
\end{equation}
Let $S_j:=\Psi_L(Y(e_j))$, where $e_j$ is the $j$th canonical vector of $\R^{n_\xi}$, such that, by real linearity,
\[
\Psi_L(Y(\xi))
    :=
        \sum_{j=1}^{n_\xi}
        \xi_j S_j,
\]
Then we can write \eqref{Eq:CasM}, with
\[
M(\hat\theta_m)
    :=
        \begin{bmatrix}
            {\frak r}(S_1 \hat G_m)
            & \cdots
            &{\frak r}(S_{n_\xi} \hat G_m)
        \end{bmatrix},
\]
\[
b(\hat\theta_m)
    :=
        {\frak r}( K_m \hat Y_{m}^{\rm ss}),
\]
such that 
\[
c(\xi,\hat\theta_m)
=
\norm{M(\hat\theta_m)\xi - b(\hat\theta_m)}_2.
\]
Hence, $M_{e,k}(\hat\theta_m)$ is simply the group of columns associated with block $\eta_{e,k}$.

\begin{pfof}{Theorem~\ref{thm:psi-frf-w1}}
Recall from Definition \ref{def:wasserstein-ambiguity-set} that for a stacked vector $\delta:=\col\big(\delta_{e,k}\big)_{(e,k)\in{\cal I}}$, 
\[
\norm{\delta}_{W}
:=
\sum_{(e,k)\in{\cal I}}
\norm{W_{e,k}\delta_{e,k}}_2.
\]
We first show that ${\cal L}(\hat\theta_m)$ from \eqref{eqn:psi-w1-block-induced-norm} is equivalently given by
\begin{equation}\label{Eq:LMax}
{\cal L}(\hat\theta_m)
    =
        \sup_{\norm{\delta}_W \le 1}
        \norm{M(\hat\theta_m)\delta}_2,
\end{equation}
i.e., is the induced $W \to 2$ norm of the matrix $M(\hat{\theta}_m)$. Since each weighting matrix $W_{e,k}$ is invertible,
\[
    \begin{aligned}
        \norm{M(\hat\theta_m)\delta}_2
        &\le
            \sum_{(e,k)\in{\cal I}}\norm{M_{e,k}(\hat\theta_m)\delta_{e,k}}_2 \\
         &=
            \sum_{(e,k)\in{\cal I}}\norm{M_{e,k}(\hat\theta_m) W_{e,k}^{-1}(W_{e,k} \delta_{e,k})}_2 \\
         &\le
            {\cal L}(\hat\theta_m)
            \sum_{(e,k)\in{\cal I}}
            \norm{W_{e,k} \delta_{e,k}}_{2} \\
        &=
            {\cal L}(\hat\theta_m)
            \norm{\delta}_{W}.
    \end{aligned}
\]
While we omit the details, standard arguments can be used to show the supremum is attained at $\mathcal{L}(\hat\theta_m)$ by constructing a vector $\delta$ concentrated at the maximum block $(e_{\star},k_{\star})$, and this establishes \eqref{Eq:LMax}.
%
%
%
Using this, now note that for $\xi,\xi'\in \Xi$ and fixed $\hat{\theta}_m$, the reverse triangle inequality gives the bound
    \begin{align*}
        \quad |
            c(\xi,\hat\theta_m)
            -
            c(\xi',\hat\theta_m)
        |
         &\leq
            \|M(\hat\theta_m)(\xi-\xi')\|_2
        \\ & \le
            {\cal L}(\hat\theta_m)\norm{\xi-\xi'}_{W},
    \end{align*}
and hence, $c(\xi,\hat\theta_m)$ is ${\cal L}(\hat\theta_m)$-Lipschitz under the weighted ground metric. Furthermore, cost $c(\xi,\hat\theta_m)$ is continuous in $\xi$, and has finite expectation under the empirical distribution $\hat{\mathbb{P}}$, and thus $c$ satisfies \cite[Assumption 1]{kuhn2024wassersteindistributionallyrobustoptimization}. The Kantorovich-Rubenstein theorem \cite[Theorem~2]{kuhn2024wassersteindistributionallyrobustoptimization} now gives, for every $\Prob\in{\cal B}_\varepsilon(\hat\Prob)$ 
    \[
    \E_\Prob[c(\xi,\hat\theta_m)]
    \le
    \E_{\hat\Prob}[c(\xi,\hat\theta_m)] 
    +
    \varepsilon_{}{\cal L}(\hat\theta_m).
    \]
    Taking the supremum proves the upper bound in Theorem \ref{thm:psi-frf-w1}. Finally, by \cite[Theorem 10]{kuhn2024wassersteindistributionallyrobustoptimization}, equality holds when  $\Xi=\R^{n_\xi}$, since $c(\xi,\hat\theta_m)$ is convex.
\end{pfof}

\begin{pfof}{Proposition \ref{Prop:BoundEstimationError}} For $e\in{\cal E}$ and $q \in \mathcal{Q}$ consider the dataset $(\hat{U}_{m}^{(e)},\hat{Y}_{m}^{(e,q)})$ with $\hat{Y}_{m}^{(e,q)} = Y_{\mathrm{sys},}^{(e,q)} + N_{m}^{(e,q)}$. By Assumption \ref{ass:repeated-experiment-model}, there is a state spectrum $\hat{X}_{m}^{(e,q)}$ such that \eqref{eqn:non-SS_IO_spectrum} holds. Let 
\[
\begin{aligned}
\bar X_m^{(e,q)} &:= \E[\hat X_m^{(e,q)}], \quad \bar Y_m^{(e,q)} := \E[\hat Y_m^{(e,q)}]\\
\overline{\Delta x}^{\,{(e,q)}} &:= \E[x_0^{(e,q)}-x_{2M}^{(e,q)}],
\end{aligned}
\]
which are all well-defined under Assumption \ref{ass:repeated-experiment-model}. Taking expectations in \eqref{eqn:non-SS_IO_spectrum} we find using Assumption \ref{ass:repeated-experiment-model} that
\begin{equation}
\label{eqn:mean-spectrum-recovery-1}
\begin{aligned}
    e^{j\hat\omega_m}\bar X_m^{(e)}
    &=
    A\bar X_m^{(e,q)} + B\hat U_m^{(e)} + e^{j\hat \omega_m}\overline{\Delta x}^{\,{(e,q)}},
\\[2ex]
    \bar Y_m^{(e,q)}
    &= Y_{\mathrm{sys},m}^{(e,q)} = 
    C \bar X_m^{(e,q)} + D \hat U_m^{(e)}.
\end{aligned}
\end{equation}
Thus, the mean data $(\hat{U}^{(e)},\bar Y_{m}^{(e,q)})$ are IO spectra of \eqref{eqn:lti-system}, and thus satisfy the equations in Corollary \ref{cor:noise-free-FRF-recovery}, which we write as
\begin{subequations}\label{Eq:ProofME}
    \begin{align}\label{Eq:ProofME-1}
            D_L \bar G_m
    &=
    \begin{bmatrix}
        W_L(z_m) \otimes \hat U_m^{({\cal E})} \\ 0
    \end{bmatrix}\\
    \label{Eq:ProofME-2}
    \bar \Psi_L\bar{G}_{m} &= K_m\bar{Y}_{m}^{\rm ss},    
    \end{align}
\end{subequations}
for a solution pair $(\bar{G}_{m},\bar{Y}_{m}^{\rm ss})$, where for compactness we have written $\bar \Psi_L = \Psi_{L}(\bar Y^{(\cal{E})})$ and written
\[
D_L:=
\begin{bmatrix}
    \Psi_L(\hat U^{(\cal{E})}) \\ \Psi_L(\hat \Omega^{(\cal{E})})
\end{bmatrix}
\]
is the constraint matrix of the deterministic constraints. By construction, it holds that
\begin{equation}
\label{eqn:FRF-expected-Y}
    H(z_m) = \bar Y_{m}^{\rm ss} (\hat U_m^{({\cal E})})^\dagger.
\end{equation}
Now let $(\hat Y^{\mathrm{ss},\star}_m,\hat G_m^\star,\hat H^{\star}(z_m))$ be defined by the estimator \eqref{eqn:optimal-frf-estimate}--\eqref{Eq:FRFEstimator}, and let
\begin{equation}\label{Eq:RandomResidual}
    r_Y^{\star}(\xi) := r_Y(\xi,\hat\theta_m^{\star})
    =
    \Psi_L\big(Y^{(\cal{E})}(\xi)\big) \hat G_m^{\star}
    -
    K_m \hat Y_{m}^{\rm ss,\star},
\end{equation}
denote the residual \eqref{Eq:DRResidual} evaluated at this optimal solution; note that $r_Y^{\star}$ is a random variable with mean
\[
\bar{r}_Y^{\star} := \mathbb{E}_{\mathbb{P}^{\star}}[r_Y(\xi,\hat{\theta}_m^{\star})]  = \bar \Psi_L \hat G_m^{\star}
    -
    K_m \hat Y_{m}^{\rm ss,\star}
\]
where we used linearity of $\Psi_L \circ Y^{(\mathcal{E})}$. Subtracting $0 =  \bar\Psi\bar{G}_{m} - K_m\bar{Y}_{m}^{\rm ss}$ from the above, we may equivalently write
\[
\bar{r}_Y^{\star} = \bar\Psi_L(\bar G_{m} - \hat G_m^{\star}) - K_m(\bar{Y}_{m}^{\rm ss} - \hat{Y}_{m}^{\rm ss,\star}).
\]
By construction, the solution component $\hat G^{\star}_{m}$ also satisfies the deterministic constraints \eqref{Eq:ProofME-1}, and thus $D_L(\bar G_{m} - \hat G_m^{\star}) = 0$. If $N_D$ is a matrix having orthonormal columns spanning $\ker D_L$, then $\bar G_{m} - \hat G_m^{\star} = N_DX$ for some matrix $X$. Therefore we may write
\[
\bar{r}_Y^{\star} = \bar\Psi_LN_DX - K_m(\bar{Y}_{m}^{\rm ss} - \hat{Y}_{m}^{\rm ss,\star}).
\]
Consider now the subspace $\operatorname{range}(\bar\Psi_LN_D)
\subseteq\C^{Ln_y}$ and its orthogonal projector
\begin{equation}
\label{eqn:frf-nuisance-projector}
P_D^\bot 
:=
I- (\bar\Psi_L N_D)(\bar\Psi_L N_D)^\dagger.
\end{equation}
It follows immediately that
\[
P_D^{\perp}K_m(\bar{Y}_{m}^{\rm ss} - \hat{Y}_{m}^{\rm ss,\star}) = P_D^{\perp}\bar{r}_{Y}^{\star}
\]
and therefore
\begin{equation}\label{Eq:IntBound}
\|P_D^{\perp}K_m(\bar{Y}_{m}^{\rm ss} - \hat{Y}_{m}^{\rm ss,\star})\|_{F} = \|P_D^{\perp}\bar{r}_{Y}^{\star}\|_{F} \leq \|\bar{r}_{Y}^{\star}\|_{F}.
\end{equation}
We can bound further \eqref{Eq:IntBound} using Jensen's inequality,  \eqref{Eq:JThetaM} and Theorem~\ref{thm:psi-frf-w1} 
\begin{equation}
\label{eqn:Jensens-bound}
\begin{aligned}
\|\bar{r}_{Y}^{\star}\|_{F}
&=
\|\E_{\Prob^\star}[r_Y^\star]\|_{F} \le
\E_{\Prob^\star}[\|r_Y^\star\|_F]
\le
\bar J^\star.
\end{aligned}
\end{equation}

We now claim there is a constant $\kappa_m > 0$ such that
\begin{equation}\label{Eq:KappaBound}
\kappa_m\|\bar{Y}_{m}^{\rm ss} - \hat{Y}_{m}^{\rm ss,\star}\|_{F} \leq  \|P_D^{\perp}K_m(\bar{Y}_{m}^{\rm ss} - \hat{Y}_{m}^{\rm ss,\star})\|_{F}.
\end{equation}
Let $\Delta Y_{m}^{\rm ss}:=\bar{Y}_{m}^{\rm ss} - \hat{Y}_{m}^{\rm ss,\star}$ and suppose by contradiction that $\kappa_m = 0$. Then there would exist  $\Delta Y_{m}^{\rm ss}\neq 0$ such that $P_D^\bot K_m \Delta Y_{m}^{\rm ss} =0$, implying that every column of $K_m \Delta Y_{m}^{\rm ss}$ belongs to $\operatorname{range}(\bar\Psi_LN_D)$. Hence, there exists a matrix $X$ such that
    \[
    K_m\Delta Y_{m}^{\rm ss} = \bar\Psi_L N_D X.
    \]
    Setting $\Delta G := N_D X$ gives
    \[
    \begin{aligned}
    D_L \Delta G &= 0,\\
    \bar\Psi_L \Delta G &= K_m \Delta Y_{m}^{\rm ss}
    \end{aligned}
    \]
    since the columns of $N_D$ span $\ker(D_L)$. Comparing these equations to \eqref{Eq:ProofME}, we conclude that $(\bar{G}_{m} + \Delta G,\bar{Y}_{m}^{\rm ss} + \Delta Y_{m}^{\rm ss})$ is also a solution of \eqref{Eq:ProofME}, which contradicts the uniqueness statement of Corollary \ref{cor:noise-free-FRF-recovery}, and hence $\kappa_m = \sigma_{\rm min}(P_{D}^{\perp}K_m) > 0$. Using \eqref{Eq:KappaBound} in \eqref{eqn:Jensens-bound}  we find that
    \begin{equation}
    \label{eqn:DeltaY-le-Jstar}
    \|\bar{Y}_{m}^{\rm ss} - \hat{Y}_{m}^{\rm ss,\star}\|_{F} 
    \leq 
    \kappa_m^{-1}\bar J^\star.
    \end{equation}
By \ref{ass:excited-input-rank}, the smallest singular value of $U_m^{({\cal E})}$ is positive for each $m\in{{\cal M}_{\rm ex}}$, which we denote  $\lambda_m := \sigma_{\rm min}(U_m^{(\mathcal{E})}) >0$ . Therefore, if $\Prob^\star \in {\cal B}_{\varepsilon}(\hat\Prob)$, then by \eqref{eqn:FRF-expected-Y} and \eqref{eqn:DeltaY-le-Jstar}
    \[
    \|{H}(z_m) - \hat{H}^{\star}(z_m)\|_{F} 
    \leq      
    \lambda_m^{-1}\kappa_m^{-1}\bar J^\star,
    \qquad
    m\in{{\cal M}_{\rm ex}}
    \]
which gives \eqref{eqn:type-1-frf-error-bound} for $\gamma_m=\lambda_m^{-1}\kappa_m^{-1}$.
\end{pfof}

\end{document}